\documentclass[]{spie}  

\usepackage{amsmath,amsfonts,amssymb}
\usepackage{graphicx}
\usepackage[colorlinks=true, allcolors=blue]{hyperref}

\title{Data downlink prioritization using image classification on-board a 6U CubeSat}

\author[a]{Keenan A. A. Chatar}
\author[a]{Ezra Fielding}
\author[a]{Kei Sano}
\author[a]{Kentaro Kitamura}
\affil[a]{Kyushu Institute of Technology, 1-1 Sensuicho, Tobata Ward, Kitakyushu City, Fukuoka Prefecture, Japan 804-0015}

\authorinfo{Further author information: (Send correspondence to Keenan Chatar)\\ E-mail: chatar.keenan-alexsei240@mail.kyutech.jp}

\usepackage{gensymb}
 
\begin{document} 
\maketitle

\begin{abstract}
Nanosatellites are proliferating as low-cost dedicated sensing systems with lean development cycles. Kyushu Institute of Technology (Kyutech) and collaborators have launched a joint venture for a nanosatellite mission, Visible Extra-galactic background RadiaTion Exploration by CubeSat (VERTECS). The primary mission is to elucidate the formation history of stars by observing the optical-wavelength cosmic background radiation. The VERTECS satellite will be equipped with a small-aperture telescope and a high-precision attitude control system to capture the cosmic data for analysis on the ground. However, nanosatellites are limited by their onboard memory resources and downlink speed capabilities. Additionally, due to a limited number of ground stations, the satellite mission will face issues meeting the required data budget for mission success. To alleviate this issue, we propose an on-orbit system to autonomously classify and then compress desirable image data for data downlink prioritization and optimization. The system comprises a prototype Camera Controller Board (CCB) which carries a Raspberry Pi Compute Module 4 which is used for classification and compression. The system uses a lightweight Convolutional Neural Network (CNN) model to classify and determine the desirability of captured image data. The model is designed to be lean and robust to reduce the computational and memory load on the satellite. The model is trained and tested on a novel star field dataset consisting of data captured by the Sloan Digital Sky Survey (SDSS). The dataset is meant to simulate the expected data produced by the 6U satellite. The compression step implements  GZip, RICE or HCOMPRESS compression, which are standards for astronomical data. Preliminary testing on the proposed CNN model results in a classification accuracy of about 100\% on the star field dataset, with compression ratios of 3.99, 5.16 and 5.43 for GZip, RICE and HCOMPRESS that were achieved on tested FITS image data
\end{abstract}


\section{Introduction}
\label{sec:intro}  

CubeSat space telescopes have emerged as significant advancements in space technology, offering cost-effective and compact solutions for astronomical observations. These miniature satellites, based on the CubeSat standard\cite{heidt2000cubesat}, provide compelling alternatives to traditional space missions due to their reduced development timelines, lower launch costs, and increased accessibility for researchers and educational institutions. CubeSat-based space telescopes have been successfully deployed in various astronomical studies, including Earth observation and exploration of celestial phenomena \cite{racca2016euclid, bowman2022cubespec, schwartz20226u}. 

The Extra-galactic Background Light (EBL) is the faint glow of electromagnetic radiation that permeates the entire universe. It consists of all light emitted by stars, galaxies, and other cosmic sources throughout cosmic history \cite{mattila2019extragalactic}. The EBL carries crucial information about the evolution and history of the universe. The sensitivity of the EBL detection relies crucially on the telescope's aperture size and its field of view, with larger apertures and wider fields of view leading to enhanced detection capabilities. In the context of CubeSat missions, a strategic approach involves employing small, yet wide-field telescopes. By opting for such compact configurations, CubeSats can effectively detect the EBL despite their limited size, presenting a unique solution to capturing the cosmic glow. 

   \begin{figure} [ht]
   \begin{center}
   \begin{tabular}{c} 
   \includegraphics[height=5cm]{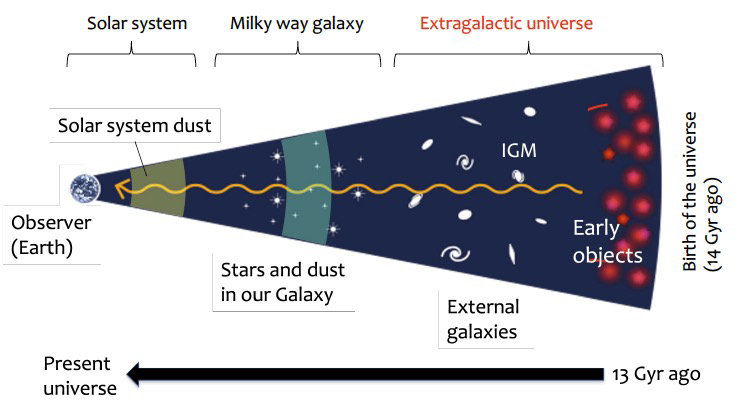}
	\end{tabular}
	\end{center}
   \caption[example] 
   { \label{fig:EBL} 
Birth of the Universe}
   \end{figure} 

Visible Extra-galactic background RadiaTion Exploration by CubeSat (VERTECS) is part of the JAXA-Small Satellite Rush Program (JAXA-SMASH), a research and development initiative fostering collaborations between universities, private companies, and JAXA to realize small satellite missions. VERTECS aims to investigate the origins of the optical-spectrum extra-galactic background radiation, contributing to our understanding of the formation and evolution of stars and galaxies. The VERTECS satellite is underway in the 2-year satellite development life-cycle and will be launched by a private Japanese rocket launch service selected by JAXA.

While CubeSats offer several advantages, they also face several limitations that can impact their performance in certain missions. One prominent disadvantage lies in their size and weight constraints. Being small and compact, CubeSats have limited payload capacity, which restricts their ability to carry more sophisticated electronics and scientific instruments. This constraint significantly affects the computational power and onboard storage capacity available for data processing and analysis during missions. Moreover, the small size of CubeSats also affects their communication capabilities. With the limited surface area available for communication systems and antennas, CubeSats may encounter challenges in maintaining strong and reliable communication links with ground stations. As a result, data transmission rates may be reduced, and the amount of data that can be sent to ground stations may be limited. This can hinder real-time data retrieval and timely analysis, particularly in missions that require rapid response or continuous monitoring\cite{palo2014expanding}.

To overcome the challenges posed by limited downlink speed in CubeSats, a promising approach is leveraging machine learning techniques for onboard data preprocessing.\cite{pritt2017onboard} By implementing machine learning algorithms directly on the satellite, the data can be intelligently analyzed, reduced, or prioritized before transmission to ground stations. These preprocessing steps allow CubeSats to efficiently manage their limited communication resources and optimize the use of downlink bandwidth. For example, machine learning algorithms can be tailored to perform data compression onboard the satellite. \cite{sculley2006compression} These algorithms can identify and discard redundant or less critical information, reducing the data size while preserving the essential scientific content. By transmitting only the most relevant data, CubeSats can minimize downlink times, facilitating faster data dissemination to researchers and decision-makers. However, for applications where data integrity is of utmost importance, such as the VERTECS mission, which entails capturing astronomical data for on-ground science analysis, we should consider traditional lossless methods of compression such as those described in Ref. \citenum{pence2009lossless} and \citenum{mandeel2021comparative}.

Additionally, onboard machine learning models can autonomously classify and label data, prioritizing valuable information for immediate transmission.  For instance, in Earth observation missions, the model can quickly identify regions of interest, such as natural disasters or environmental changes, enabling real-time alerts and timely responses.\cite{thompson2015onboard, azami2022earth, gretok2021onboard} This technique can be tailored for astronomical observation to label and identify regions of interest or discard/recapture invalid data. By intelligently selecting crucial data for downlink, CubeSats can effectively address the challenges posed by limited communication capabilities, maximizing the scientific output of their missions and advancing our understanding of the universe. As machine learning algorithms continue to advance and become more efficient, their integration into CubeSat operations holds great promise for overcoming the limitations of downlink speed and enhancing the overall effectiveness of small satellite missions.

\section{Mission Objectives}
\label{sec:mission_objectives}
\begin{figure} [ht]
   \begin{center}
   \begin{tabular}{c} 
   \includegraphics[height=7cm]{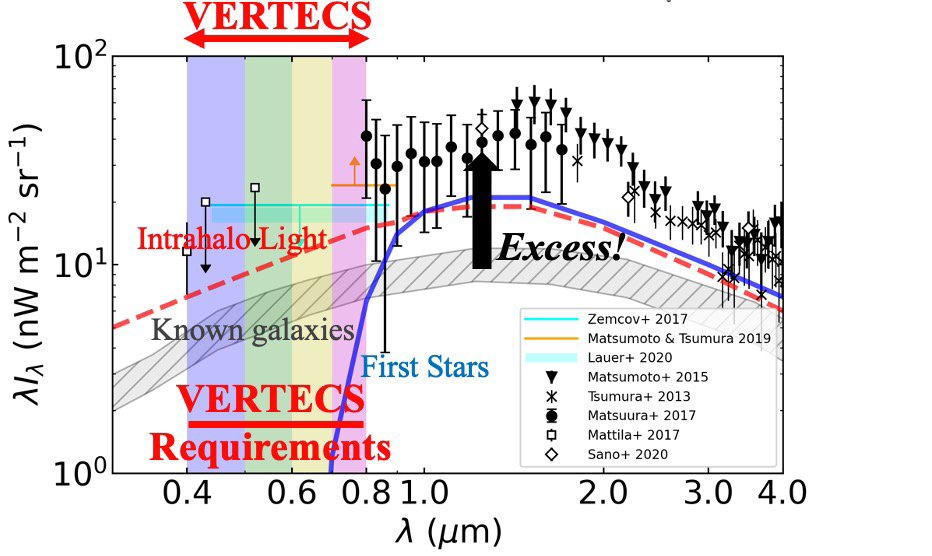}
	\end{tabular}
	\end{center}
   \caption[example] 
   { \label{fig:EBL_study} 
VERTECS Observation Range and previous EBL Measurements}
   \end{figure}

Studies of the near-infrared Extra-galactic Background Light (EBL) have revealed that its brightness surpasses the combined integrated light emitted by known galaxies. \cite{sano2020isotropic, windhorst2022skysurf, o2022skysurf} Figure~\ref{fig:EBL_study} illustrates the surface brightness of the previous EBL measurements and the measurement wavelength. The discrepancy between the measured EBL brightness and the expected contribution from known galaxies suggests the existence of unknown faint extra-galactic sources. 

One possible explanation for this phenomenon is the presence of Intrahalo Light (IHL) which refers to the faint glow of light emanating from stars distributed in the outskirts of galaxy halos \cite{cooray2012near}. The cumulative light from these dispersed stellar populations can become significant, accounting for the excess near-infrared EBL brightness observed in studies. Another possibility is the potential contribution of the First Stars in the universe's early stages. The First Stars, also known as Population III stars, are believed to have formed from pristine hydrogen and helium shortly after the Big Bang \cite{matsumoto2019fluctuation}. These stars are hypothesized to be massive and short-lived, emitting intense radiation during their brief lifetimes. However, direct observational evidence of these ancient stars remains elusive, with only theoretical models and indirect signatures guiding our understanding of their existence. The presence of such First Stars, if confirmed, could significantly contribute to the near-infrared EBL brightness. The primary objective of VERTECS is to conduct spectral observations in the visible wavelength (0.4 $\mu$m – 0.8$\mu$m), as seen in Figure~\ref{fig:EBL_study}, to distinguish the spectrum of intra-halo light and first stellar objects. To achieve this objective, the satellite is required to capture multiple band wavelengths with high sensitivity, throughput (S$\Omega> 10^{-6} m^2 sr$), and noise rejection. 

This mission requirement leads to an extensive observation requirement per day, and subsequently a large data downlink budget. The projected data downlink requirement is approximately 608 MB of data per day. Nanosatellites typically face limitations in terms of onboard memory resources and downlink speed capabilities. Furthermore, limited access to ground stations poses challenges to meeting the necessary data budget for mission success. To address this issue, we propose an on-orbit data processing pipeline to autonomously classify, prioritize and compress captured data onboard the satellite before downlinking. The system is designed to first classify images, followed by either compressing preferred images for later downlink or discarding/recapturing invalid data. The system uses a machine-learning model that is trained on a novel star field dataset to classify image data onboard the satellite. 

\section{System Design}
\label{sec:sys_des}
\subsection{Satellite Design}

The VERTECS CubeSat is designed using the 6U CubeSat standard and primarily consists of the Imaging Telescope, AOCS System, and Bus/Payload System. The Main Bus System is developed by the Kyushu Institute of Technology (Kyutech) and is based on the BIRDS Bus configuration. \cite{kim2021birds} The primary components of the BIRDS Bus are the Onboard Computer (OBC), the Electrical Power System (EPS), the Communication System (COM), and the Back-plane Board (BPB). The OBC comprises a series of PIC microcontrollers and is the primary command and data-handling subsystem that coordinates the operation of the other subsystems and payload interface boards. The EPS comprises a microcontroller, battery assembly, and solar panel assembly. This system has the vital role of providing a constant steady power supply for all onboard electronics and the mission payload during daytime and eclipse. Two communication subsystems will be installed on the VERTECS satellite. First, an S-Band communication system that interfaces directly to the OBC will be the primary method for command uplink and telemetry data downlink. An additional X-band communication assembly that interfaces directly with the payload interface electronics will be mainly used for downloading mission data to the ground stations.

   \begin{figure} [ht]
   \begin{center}
   \begin{tabular}{c} 
   \includegraphics[height=7cm]{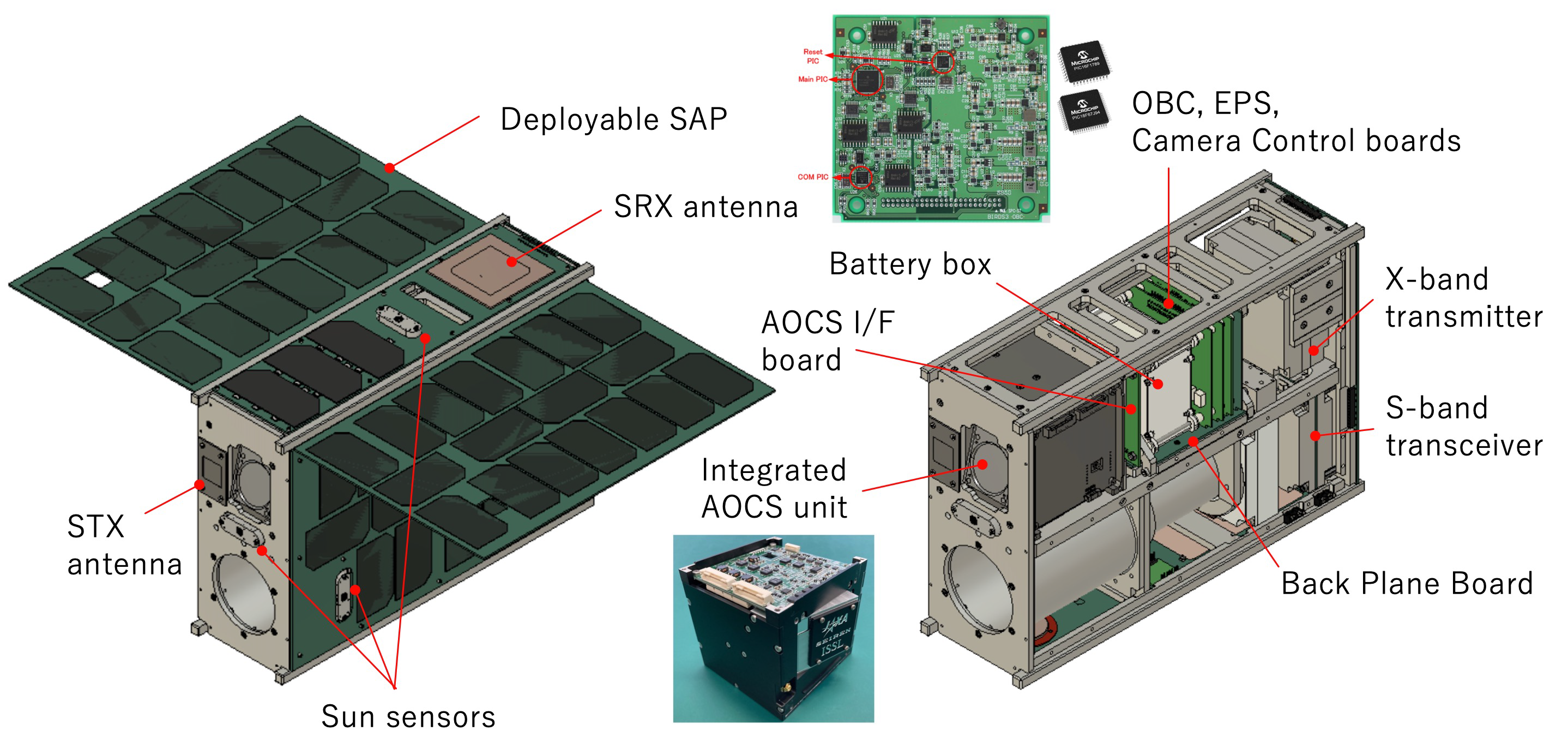}
	\end{tabular}
	\end{center}
   \caption[example] 
   { \label{fig:VERTECS_system_design} 
VERTECS Satellite Design}
   \end{figure} 

\subsection{Attitude and Orbit Control System}
VERTECS will be equipped with a 1U Attitude and Orbit Control System (AOCS) aimed at actively controlling the satellite's orientation during missions and executing various maneuvers, including B-dot stabilization mode, Sun tracking mode, Nadir pointing mode, and Inertial pointing mode. The AOCS comprises essential hardware components, such as a Control unit, three Reaction Wheels (RW), three Magnetorquers, one gyroscope, one magnetometer, one star tracker, and one General Packet Radio Service (GPRS) module. Additionally, an AOCS adapter board facilitates interfacing with the satellite's systems, while an external Magnetometer and six sun sensors enhance its capabilities. The integrated hardware functions in concert to enable active control over the satellite's orientation, critical for mission success and specific objectives. The B-dot mode utilizes magnetic field-based attitude control, the Sun tracking mode optimizes solar panel orientation, Nadir pointing mode aligns the satellite's antennas toward Earth's surface for optimum communication, and the Inertial pointing mode ensures accurate orientation relative to the stars for precise targeting.

Beyond attitude control, the AOCS also performs functions such as Orbit propagation and determination, Target attitude calculation, FDIR (Fault Detection, Isolation, and Recovery) function for fault management, and Simulation function for testing and verification. The system also includes communication functions for interaction with the satellite's bus system, Data Storage for onboard retention, power switching and monitoring, and over-current protection mechanisms for operational stability. Moreover, the AOCS incorporates a Temperature measurement function to monitor and regulate the satellite's thermal conditions effectively. With its diverse set of hardware components and advanced control algorithms, this Attitude and Orbit Control System will empower the VERTECS satellite to adapt flexibly to varying mission requirements and maintain stable and accurate attitude control throughout its operations in space, ensuring optimal performance and successful mission accomplishment.

\subsection{Imaging Payload}
VERTECS will carry a 2U imaging payload, incorporating a wide-field optical telescope assembly and its corresponding control/interface electronics. The primary objective of this system is to accurately measure the spectral energy distribution within visible wavelengths. The lens optics are designed to achieve high-throughput capabilities, with solid angle, $S\Omega$ surpassing $10^{-6} \, \text{m}^2 \, \text{sr}$, fulfilling the mission requirements. The system's implementation encompasses a lens diameter (D) of 3.5 cm and a broad field of view (FoV) spanning 6 degrees. To optimize overall performance, the imaging payload is equipped with a baffle system, rejecting stray light arising from both the Sun and Earth. The stray-light rejection is governed by the gain function \( g(\theta) \), and the system ensures that the gain function ratio \( \frac{g(0^\circ)}{g(30^\circ)} \) exceeds a value of \( 10^6 \).

Moreover, the payload integrates a color filter, enabling multi-band imaging capabilities across the optical region, encompassing wavelengths from 400 nm to 800 nm. This versatile feature empowers the system to capture data in four distinct bands, significantly enhancing its scientific observation and research potential. The satellite aims to achieve a multi-color image of a 3-degree square area by overlapping the 1/4-field shifted images through four observations in a single orbit. The observation area selection conditions require a solar elongation of greater than 90° and an Earth elongation of approximately 180°. Each image captures a 60-second exposure time, and the system maintains an absolute pointing accuracy of less than 0.01 deg (3$\sigma$). For mapping a broader area, the system tiles the images during multiple orbits, providing comprehensive coverage and expanding its scientific capabilities.

   \begin{figure} [ht]
   \begin{center}
   \begin{tabular}{c} 
   \includegraphics[height=9cm]{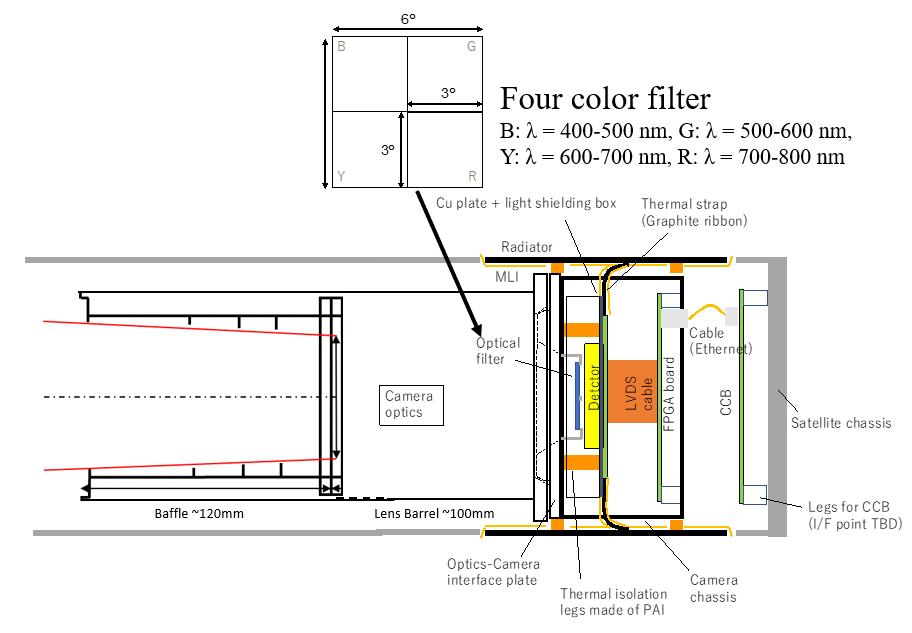}
	\end{tabular}
	\end{center}
   \caption[example] 
   { \label{fig:VERTECS_telescope} 
VERTECS Imaging Payload Assembly}
   \end{figure} 

\subsection{Camera Controller Board}
The VERTECS Camera Control Board (CCB) will be the primary interface board between the mission payload and the BIRDS BUS system. The CCB plays a critical role in managing various aspects of the satellite mission. It is responsible for executing mission commands, overseeing data collection and storage, and facilitating data transfer. Additionally, the CCB provides essential interface connections with the mission payload, enabling efficient command execution. It also interfaces with the X-band Transmitter, enabling the downlink of mission data. Moreover, the CCB is designed to accommodate the satellite's physical limitations, ensuring its resilience in the space environment and enabling smooth operation throughout the expected lifetime of the satellite.

The primary microcontroller is the Raspberry Pi Compute Module 4 (CM4) with 8GB of RAM and 32GB of embedded flash memory storage. This module was chosen due to its low cost, availability, and robust performance. The CM4 communicates with the OBC via UART for command and housekeeping data transfer. The primary interface between the CCB and the payload Camera FPGA will be a standard Ethernet connection. The CM4 runs Ubuntu Server 22.04.02 LTS and the Ethernet functionality built into the device is used. The CCB will also directly interface with the X-band communication system via a high-speed UART connection for high-speed data transfer and downlink. The CCB hosts all code and scripts used for the prioritization pipeline written in Python. Additional libraries are installed to implement the classification algorithm and compression methods which are described in the next section. The CCB is still under active design. The CM4 mounted to a Raspberry Pi Compute Module 4 IO board will be used to test functionality, interface connections and performance while the CCB design is finalized.

\begin{table}[h]
\centering
\caption{CM4 Specifications} 
\label{tab:CM4-Specifications}
\begin{tabular}{|l|l|}
\hline
\textbf{Parameters} & \multicolumn{1}{c|}{\textbf{Details}}                                \\ \hline
Processor           & Broadcom BCM2711, Quad-core Cortex-A72 (ARM v8) 64-bit SoC, @1.5 GHz \\ \hline
RAM                 & 8GB LPDDR4-3200 SDRAM.                                               \\ \hline
Storage             & 32GB Embedded Flash                                                  \\ \hline
Ethernet            & Gigabit Ethernet                                                     \\ \hline
Power               & $\sim$1.2W Idle @5V DC Supply                                        \\ \hline
\end{tabular}
\end{table}

\begin{figure} [ht]
   \begin{center}
   \begin{tabular}{c} 
   \includegraphics[height=9cm]{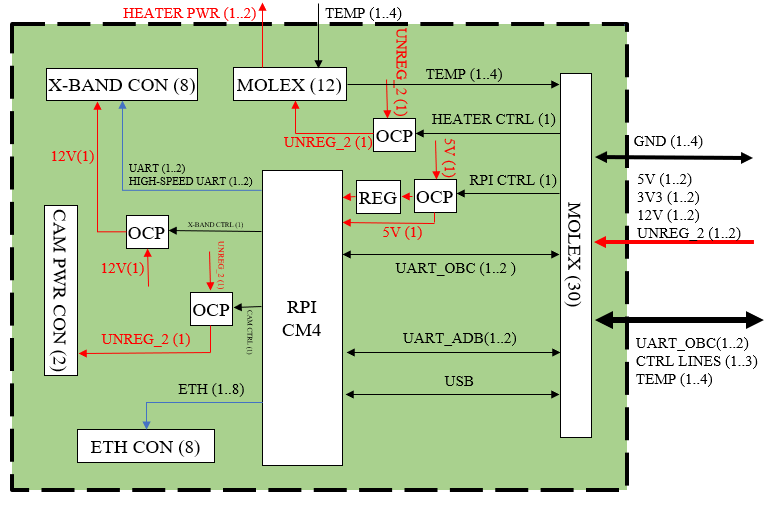}
	\end{tabular}
	\end{center}
   \caption[example] 
   { \label{fig:CCB_block_diagram} 
CCB Block Diagram}
   \end{figure}

\section{Image Prioritization System}
\label{sec:prepro_sys}

\subsection{Prioritization Pipeline}

For this work, an image prioritization pipeline is developed for on-orbit data classification, prioritization, and compression to improve data downlink efficiency. This pipeline primarily comprises a classification step and a compression step. The CubeSatNet \cite{maskey2020cubesatnet}  Convolutional Neural Network (CNN) architecture was selected as the base architecture for the CNN implemented in this pipeline. The classifications will be used to prioritize desired images for downlink. The compression methods used in this work are GZip\cite{gailly1992gnu}, RICE\cite{rice1993algorithms} and HCOMPRESS\cite{postman1992_HCOMPRESS} file compression in the AstroPy Python library\cite{price2022astropy}. These compression methods are widely used in the astronomical community \cite{pence2009lossless} as it significantly reduces the data size without sacrificing crucial information, ensuring optimal utilization of onboard memory and expedited data transmission during satellite missions and other data-intensive tasks.

\begin{figure} [ht]
   \begin{center}
   \begin{tabular}{c} 
   \includegraphics[height=8.5cm]{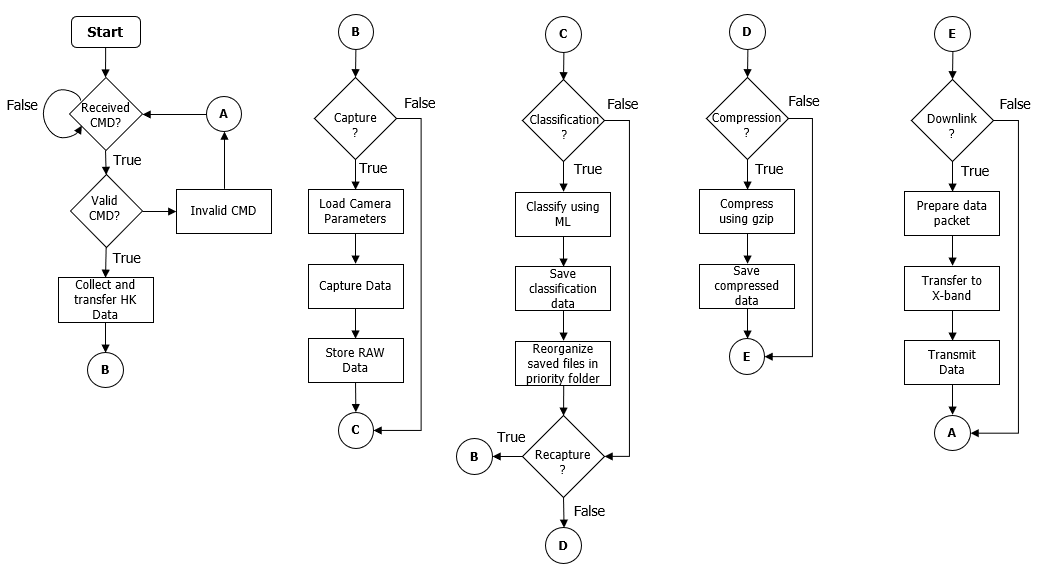}
	\end{tabular}
	\end{center}
   \caption[example] 
   { \label{fig:Preprocessing_pipeline} 
Preprocessing Flow Diagram}
   \end{figure} 

\subsection{Classification Algorithm}
 The CubeSatNet CNN architecture was chosen due to its proven flexibility and superior performance in classification tasks compared to other methods \cite{maskey2020cubesatnet}. The emphasis during the design phase was to ensure the model's lightweight nature without compromising classification accuracy. The model architecture was designed to obtain a balance between computational load and memory consumption. The input and final layers of the CubeSatNet CNN have been adapted for use in this pipeline.

Table~\ref{tab:Model-Architecture} gives an overview of the model architecture. The chosen algorithm incorporates an 11-layer design with 4 convolutional layers responsible for feature extraction. The input layer is optimized to process and resize images to a standardized 512x512 input, enabling the mission to capture images of varying resolutions and achieve precise classification results. Within the convolutional layers, the integration of Max Pooling layers effectively reduces the trainable parameters, thereby lowering the computational load during training. Finally, the model employs Global Average Pooling and a Dense function with SoftMax Activation to produce the Output Layer. This layer generates a feature map corresponding to each classification category, accompanied by confidence probabilities. This model was built using Tensorflow 2.9.1 (TF) and written in Python 3.10.9. By adopting this CNN architecture, the prioritization pipeline will achieve accurate and efficient classification, with the ability to reorganize and prioritize data based on the classification. 

\begin{table}[ht]
\centering
\caption{Model Architecture} 
\label{tab:Model-Architecture}
\begin{tabular}{|c|c|}
\hline
\textbf{Layer Type}       & \textbf{Output Shape} \\ \hline
Input                     & (512, 512, 3)         \\ \hline
Convolution + MaxPool I   & (256, 256, 16)        \\ \hline
Convolution + MaxPool II  & (128, 128, 32)        \\ \hline
Convolution + MaxPool III & (64, 64, 64)          \\ \hline
Convolution + MaxPool IV  & (32, 32, 128)         \\ \hline
Global Avg, Pooling       & (128, 1)              \\ \hline
Output Layer              & (5, 1)                \\ \hline
\end{tabular}
\end{table}

\subsection{Classification Dataset}
\begin{figure} [hb]
   \begin{center}
   \begin{tabular}{c} 
   \includegraphics[height=5.2cm]{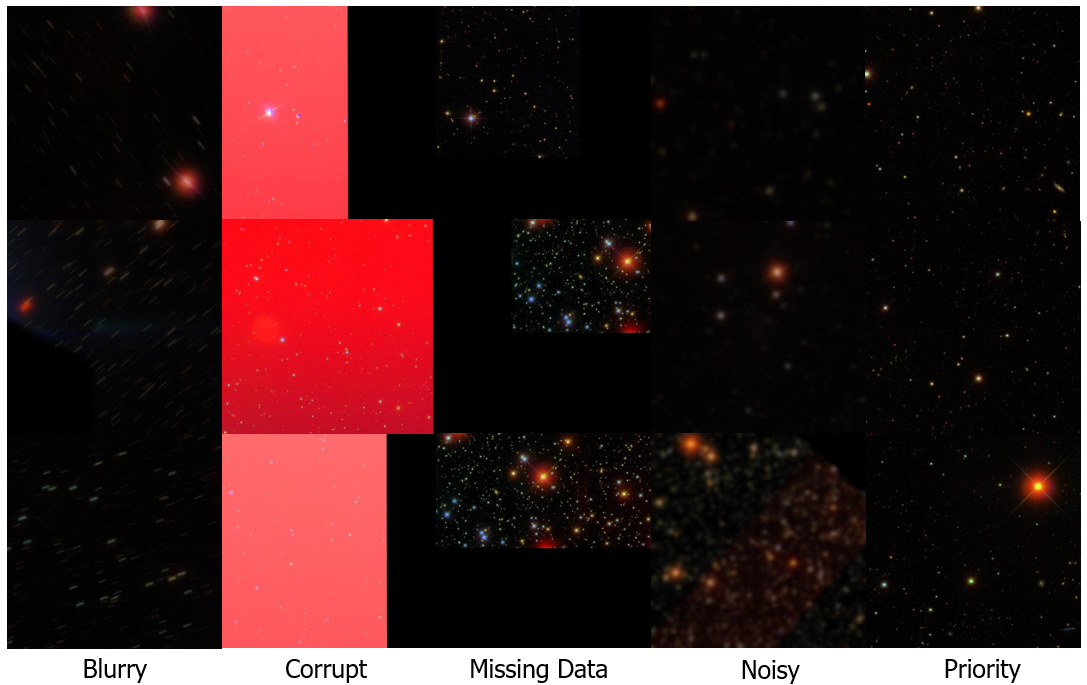}
	\end{tabular}
	\end{center}
   \caption[example] 
   { \label{fig:Dataset} 
Examples of data in VERTECS Dataset}
   \end{figure}

A novel dataset was compiled to train the CNN model to distinguish data that is considered to be desirable for scientific research. These images were compiled primarily from the Sloan Digital Sky Survey (SDSS) Data Release 16\cite{kollmeier2017sdss}. The images were selected to simulate the expected data to be captured by the VERTECS satellite. The downloaded images were subjected to rotations, distortions, reflections, and various other augment techniques to expand the size of the dataset and improve the robustness of the training data to prevent over-fitting during training. The dataset contains 5 classes to distinguish the captured data. These classes are Blurry, Corrupt, Missing Data, Noisy, and Priority. The dataset classes are designed to encompass expected erroneous data that may be captured by VERTECS during operation. Blurry data is an example of data captured as the satellite is in motion during observation sequence maneuvers. Corrupt data can occur during capture when the camera was not properly primed or the image contains defects due to stray light. Missing Data represents a partial or full loss of data in the image. Noisy data simulates data that is over-saturated with noise during capture due to radiation or other sources. Finally, Priority images represent data that is determined to be a clear image likely to be used on the ground for scientific purposes. With these classifications, the satellite can autonomously prioritize, recapture or discard data. The dataset contains a total of 20,234 images. A standard 80:20 train-test split was applied before training to create a train and test dataset. The test dataset also served as a validation dataset. Table~\ref{tab:Dataset_Support} presents the number of images present in the dataset for training and testing for each class label. Fig~\ref{fig:Dataset} displays a small subset of images found in each class.

    \begin{table}[h]
    \caption{Dataset Label Support} 
    \label{tab:Dataset_Support}
    \centering
    \begin{tabular}{|lccc|}
    \hline
    \multicolumn{1}{|c|}{\textbf{Label}} & \multicolumn{1}{c|}{\textbf{Train}} & \multicolumn{1}{c|}{\textbf{Test}} & \textbf{Total} \\ \hline
    \multicolumn{1}{|l|}{Blurry} & \multicolumn{1}{c|}{3544} & \multicolumn{1}{c|}{887} & \textbf{4431} \\ \hline
    \multicolumn{1}{|l|}{Corrupt} & \multicolumn{1}{c|}{1070} & \multicolumn{1}{c|}{268} & \textbf{1338} \\ \hline
    \multicolumn{1}{|l|}{Missing Data} & \multicolumn{1}{c|}{2021} & \multicolumn{1}{c|}{506} & \textbf{2527} \\ \hline
    \multicolumn{1}{|l|}{Noisy} & \multicolumn{1}{c|}{3582} & \multicolumn{1}{c|}{896} & \textbf{4478} \\ \hline
    \multicolumn{1}{|l|}{Priority} & \multicolumn{1}{c|}{5968} & \multicolumn{1}{c|}{1492} & \textbf{7460} \\ \hline
    \end{tabular}
    \end{table}

\section{Results and Discussion}
\label{sec:Results}
\subsection{Priority Classification}
Training of the classification algorithm was completed on a workstation PC equipped with an NVIDIA GeForce RTX 3070 Ti GPU. The training process consisted of 10 epochs, completed within 12 minutes. A batch size of 64 was chosen to align with the memory limitations of the training hardware.

\begin{figure} [h]
   \begin{center}
   \begin{tabular}{c} 
   \includegraphics[height=9cm]{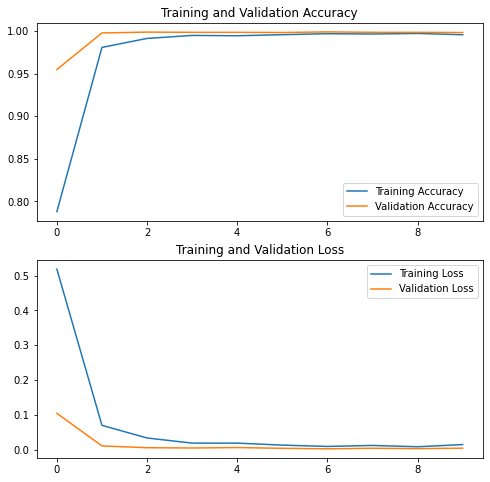}
	\end{tabular}
	\end{center}
   \caption[example] 
   { \label{fig:Classification_training} 
Classification Training and Validation Accuracy}
   \end{figure}

\begin{table}[h]
\caption{Classification Report} 
\label{tab:Classification-Report}
\centering
\begin{tabular}{|lccc|}
\hline
\multicolumn{1}{|c|}{\textbf{Labels}}           & \multicolumn{1}{c|}{\textbf{Precision}} & \multicolumn{1}{c|}{\textbf{Recall}} & \textbf{F1-Score} \\ \hline
\multicolumn{1}{|l|}{Blurry}                    & \multicolumn{1}{c|}{1.00}               & \multicolumn{1}{c|}{0.99}            & 1.00              \\ \hline
\multicolumn{1}{|l|}{Corrupt}                   & \multicolumn{1}{c|}{1.00}               & \multicolumn{1}{c|}{1.00}            & 1.00              \\ \hline
\multicolumn{1}{|l|}{Missing Data}              & \multicolumn{1}{c|}{1.00}               & \multicolumn{1}{c|}{1.00}            & 1.00              \\ \hline
\multicolumn{1}{|l|}{Noisy}                     & \multicolumn{1}{c|}{0.99}               & \multicolumn{1}{c|}{1.00}            & 1.00              \\ \hline
\multicolumn{1}{|l|}{Priority}                  & \multicolumn{1}{c|}{1.00}               & \multicolumn{1}{c|}{1.00}            & 1.00              \\ \hline
                                                &                                         &                                      &                   \\ \hline
\multicolumn{1}{|r|}{\textbf{Accuracy}}         & \multicolumn{2}{c|}{}                                                          & 1.00              \\ \hline
\multicolumn{1}{|r|}{\textbf{Macro Average}}    & \multicolumn{1}{c|}{1.00}               & \multicolumn{1}{c|}{1.00}            & 1.00              \\ \hline
\multicolumn{1}{|r|}{\textbf{Weighted Average}} & \multicolumn{1}{c|}{1.00}               & \multicolumn{1}{c|}{1.00}            & 1.00              \\ \hline
\end{tabular}
\end{table}

Fig~\ref{fig:Classification_training} presents the training and validation accuracy, as well as the training and validation loss. In both cases, the graphs relating to training and validation appear to successfully converge, indicating that the model was successfully trained within the 10 epochs. The trained model was converted from a TF Saved Model to a TFlite model. The resulting size of the TFlite model was 390KB. The TFlite model was then transferred to the Raspberry Pi Compute Module 4 for performance evaluation on the test dataset. 

The resulting discrete classifications were used to produce accuracy metrics. These metrics include precision, recall, and F1-score. Precision represents the ratio of correctly classified images belonging to the positive class to the total number of images classified in the positive class. On the other hand, recall refers to the ratio of correctly classified images belonging to the positive class to the total number of images in the positive class. The F1-score, as the weighted average of recall and precision, provides a comprehensive evaluation of the model's performance. To account for label imbalance, the Weighted Average metrics were considered. The Weighted Average considers the average precision, recall, and F1-score, weighted by the support for each class. On the other hand, the Macro Average metrics determine the average precision, recall, and F1 score across all classes without considering class imbalances.

Table~\ref{tab:Classification-Report}, presents the accuracy metrics produced after using the TFlite model to make predictions on the test dataset on the Raspberry Pi Compute Module 4. The model achieved perfect accuracy in almost every class. It should be noted that the accuracy for the Priority class is 100\% for this dataset, which is the ideal case for the proposed prioritisation pipeline. Further investigation is required to determine whether this performance scales to larger and more varied datasets. The performance of the TFlite model matches the TF model which indicates that there was no loss in performance during the conversion of the model. The Raspberry Pi Compute Module 4 completed the predictions of all 4032 images in the test dataset in 1168.74 seconds. On average, one prediction is completed in about 0.29 seconds.

\begin{figure} [hb]
   \begin{center}
   \begin{tabular}{c} 
   \includegraphics[height=11.4cm]{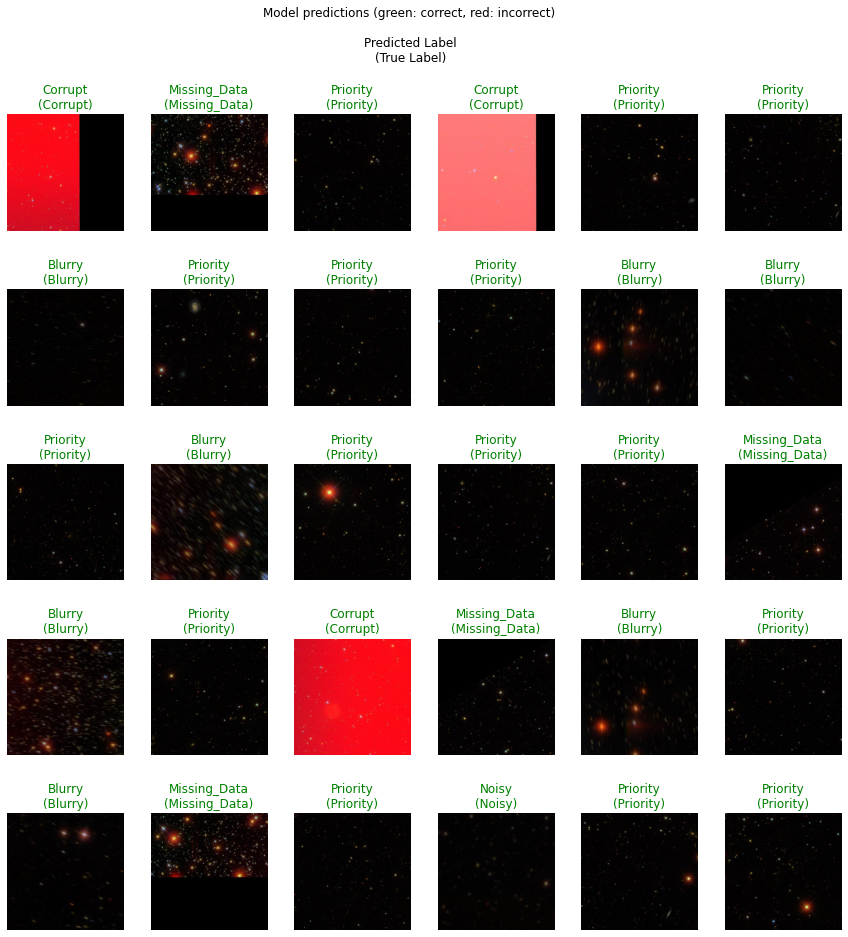}
	\end{tabular}
	\end{center}
   \caption[example] 
   { \label{fig:Classification_results} 
Classification Results}
   \end{figure}

Further testing was conducted on the validation dataset to gauge the performance of the model. A sample of 30 images is displayed in Fig \ref{fig:Classification_results} which illustrates the predictions of the algorithm for the test data. For each image, the predicted label is displayed and the ground truth label is included below in parentheses. From this data, we can see that the algorithm accurately classified each image with 100\% accuracy. In this test, the algorithm prioritizes 15 of the images as the highest priority images to download, followed by the images with missing data, corrupt, noisy, or blurry. In this simulated test set, the satellite can choose to recapture the observation sequences that were classified as irregular, or keep the data and inform ground station operators it is likely to be unusable. This is the first step towards a level of autonomy onboard a satellite that can greatly improve  in-orbit operations and efficiency. 

\subsection{Compression Study}

\begin{figure} [th]
   \begin{center}
   \begin{tabular}{c} 
   \includegraphics[height=5.8cm]{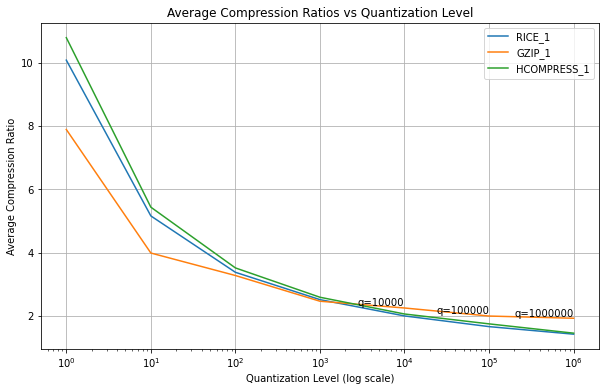}
	\end{tabular}
	\end{center}
   \caption[example] 
   { \label{fig:compression_ratio_vs_qlvl_logscale} 
Compression Ratio vs Quantization level}
   \end{figure}

\begin{figure} [th]
   \begin{center}
   \begin{tabular}{c} 
   \includegraphics[height=6.3cm]{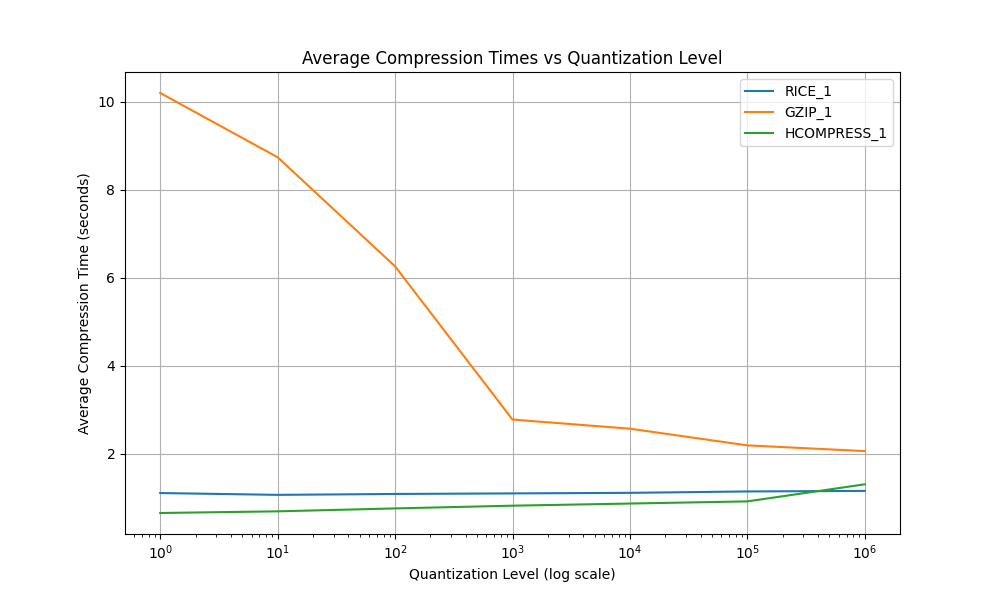}
	\end{tabular}
	\end{center}
   \caption[example] 
   { \label{fig:Figure_1_times} 
Compression times vs Quantization level}
   \end{figure}

After the classification step, the data prioritized for downlink can be compressed to reduce the download time for each observation sequence. For this work, GZip, RICE and HCompress compression algorithms are considered, and the performance is compared. Fig \ref{fig:compression_ratio_vs_qlvl_logscale} compares the compression ratio of the three algorithms at different quantization levels. Quantization is a fundamental process in digital signal processing and data compression that involves reducing the precision of data values while representing them digitally. In the context of image or data compression, quantization is used to reduce the number of distinct values that a dataset contains, thereby reducing its overall size and potentially leading to compression\cite{gray1998quantization}. Quantization level refers to the number of discrete levels into which the data values are mapped during quantization. In simpler terms, it determines the granularity of the representation. A higher quantization level indicates more discrete levels, while a lower quantization level implies fewer levels. 

From the results, we can see that at low q-levels, HCompress and RICE algorithms perform significantly better than GZip, which makes these two algorithms strong candidates for compression where preservation of detail is not the primary concern. As the q-levels increase, the compression ratios decrease for all three methods until all three methods have approximately the same compression ratios around q-level = 10000. At higher levels, GZip shows better compression ratios compared to the other algorithms, which makes GZip the best algorithm for compression when maintaining data integrity is paramount. 

Fig \ref{fig:Figure_1_times} illustrates the average compression times for the three algorithms when tested on the target hardware. At q$<$10, the RICE and HCOMPRESS algorithms perform the best with compression times around 1.17 and 1.04 seconds respectively, while GZip is significantly slower at 8.44 seconds. At higher q-levels, the compression times of GZip significantly reduce but RICE and HCOMPRESS maintain similar compression times at lower levels. At very high q-levels, GZip reduces to around 1.99 seconds, while RICE and HCOMPRESS take 1.06 and 1.2 seconds respectively. From this data, we can see that GZip compress time is the slowest at all stages, while RICE and HCOMPRESS are similar in compression speed. This information is useful when considering the computational load and power restrictions for the mission. 

Further testing was conducted to compare the performance of the fully lossless GZip compression with zero quantization, which is illustrated in Table \ref{tab:Compression-Performance}. From these results, we can see that the lossless GZip has the lowest compression ratio and compression speed when compared to all other methods, however when considering data integrity for the mission, this method will be invaluable for ensuring no data is lost during the implementation of this image prioritization and compression system. The choice of compression method and quantization level should be carefully considered to strike a balance between data reduction and computational resources during satellite operations. Effective data compression can significantly enhance the efficiency of data transfer, making it a critical aspect of the overall data management strategy for satellite missions.

\begin{table}[t]
\centering
\caption{Compression Performance} 
\label{tab:Compression-Performance}
\begin{tabular}{|c|c|c|c|c|}
\hline
\textbf{Parameter}               & \textbf{Lossless GZip} & \textbf{GZip} & \multicolumn{1}{l|}{\textbf{RICE}} & \multicolumn{1}{l|}{\textbf{HCOMPRESS}} \\ \hline
Compression Ratio                & 2.03                   & 3.99          & 5.16                               & 5.43                                    \\ \hline
Compression CPU time (seconds)   & 10.2                   & 8.44          & 1.17                               & 1.04                                    \\ \hline
Decompression CPU time (seconds) & 0.33                   & 0.35          & 0.30                               & 0.26                                    \\ \hline
\end{tabular}
\end{table}

\section{Conclusion}
\label{sec:Conclusion}
In this work, the VERTECS satellite is introduced, which is currently in the design and development phase. The objective of this satellite is to investigate the origins of the optical-spectrum extragalactic background radiation. This satellite will integrate many components for mission success and represents an advancement in space technology, harnessing the capabilities of nanosatellites for astronomical observations. 

Moreover, testing is conducted for a Camera Controller Board (CCB) equipped with a lightweight Convolutional Neural Network (CNN) classification system to improve data management on the satellite. The CNN-based classification system efficiently identifies and prioritizes desirable image data, thereby mitigating the challenges posed by limited onboard memory and downlink speed capabilities with a current test accuracy of about 100\% with an F1 Score of 1.00. Furthermore, the utilization of GZip, RICE or HCOMPRESS for data compression can optimize data transfer efficiency with a compression rate of 3.99, 5.16 and 5.43 respectively, reducing the expected data downlink time for each observation sequence while preserving essential information.

VERTECS aims to be an example of how space technology continues to evolve, enabling new opportunities for space-based research and advancing our understanding of the cosmos. The implementation of the classification and compression pipeline showcases the adaptability and efficiency in handling data-intensive missions. VERTECS is scheduled to conclude development and delivery in late 2024.

\acknowledgments 
 
The authors extend their gratitude to the entire VERTECS team for their invaluable guidance and assistance. VERTECS is sponsored by the JAXA-SMASH (JAXA-Small Satellite Rush) initiative. Initiated in December 2022, this program requires a two-year timeline for satellite development.

The JAXA-SMASH Program represents a pioneering research and development effort that fosters collaboration between universities, private enterprises, and JAXA. The program's mission is to realize small satellite missions through strategic utilization of commercial small launch opportunities. It also aims to enhance transportation services diversity within Japan.

\bibliography{report} 
\bibliographystyle{spiebib} 

\end{document}